\documentclass[prc,aps,twocolumn,showpacs,amssymb,superscriptaddress,float]{revtex4}

\usepackage{amsmath}
\usepackage{graphicx}
\usepackage{dcolumn}
\usepackage{float}

\begin{document}

\title{Short-range correlations and shell structure of medium-mass nuclei}

\author{L. Coraggio}
\affiliation{Istituto Nazionale di Fisica Nucleare, \\
Complesso Universitario di Monte  S. Angelo, Via Cintia - I-80126 Napoli,
Italy}
\author{A. Covello}
\affiliation{Istituto Nazionale di Fisica Nucleare, \\
Complesso Universitario di Monte  S. Angelo, Via Cintia - I-80126 Napoli,
Italy}
\affiliation{Dipartimento di Scienze Fisiche, Universit\`a
di Napoli Federico II, \\
Complesso Universitario di Monte  S. Angelo, Via Cintia - I-80126 Napoli,
Italy}
\author{A. Gargano}
\affiliation{Istituto Nazionale di Fisica Nucleare, \\
Complesso Universitario di Monte  S. Angelo, Via Cintia - I-80126 Napoli,
Italy}
\author{N. Itaco}
\affiliation{Istituto Nazionale di Fisica Nucleare, \\
Complesso Universitario di Monte  S. Angelo, Via Cintia - I-80126 Napoli,
Italy}
\affiliation{Dipartimento di Scienze Fisiche, Universit\`a
di Napoli Federico II, \\
Complesso Universitario di Monte  S. Angelo, Via Cintia - I-80126 Napoli,
Italy}
\author{T. T. S. Kuo}
\affiliation{Department of Physics, SUNY, Stony Brook, New York 11794,
USA}

\date{\today}

\begin{abstract}
The single-particle spectrum of the two nuclei  $^{133}$Sb and
$^{101}$Sn is studied within the framework of the time-dependent
degenerate linked-diagram perturbation theory starting from a class of
onshell-equivalent realistic nucleon-nucleon potentials.
These potentials are derived from the CD-Bonn interaction by using the
so-called $V_{\rm low-k}$ approach with various cutoff momenta.
The results obtained evidence the crucial role of short-range
correlations in producing the correct $2s1d0g0h$ shell structure.
\end{abstract}


\maketitle

\section{Introduction}
Shell structure plays a key role in explaining many features of atomic
nuclei. 
In particular, the energy gap between  groups of levels is crucial in
determining their stability properties. 
However, what is the  microscopic origin of the shell structure is
still a subject of great theoretical interest.

In recent years special attention has been focused on the role of the
various components of the nuclear force in the formation and evolution
of the shell structure.
This is mainly related to the advent of new advanced techniques and
facilities, and in particular to the development of radioactive ion
beams, which allow to study how the properties of nuclei may change
when going toward the proton or neutron drip line.
It is for instance a current matter of investigation the influence of
the tensor force on the single-nucleon energies, especially in the
case of exotic nuclei \cite{Otsuka05,Otsuka06,ABrown06}, where it is
expected to produce a reduction of the spin-orbit splitting
\cite{Zalewski08}.
This aspect is also discussed in Ref. \cite{Dobaczewski07} along with
the influence of many-body correlations and of states unbound to
particle emission on the single-particle spectrum of neutron-rich
nuclei.

In this work, we have studied the structure of the $2s1d0g0h$-shell
for protons and neutrons, calculating the single-particle spectrum of
$^{133}$Sb and $^{101}$Sn with realistic nucleon-nucleon ($NN$)
potentials. 
Both these nuclei are far from stability, $^{133}$Sb having a large
neutron excess and $^{101}$Sn lying very close to the proton drip
line.
On the other hand, $^{132}$Sn shows strong shell closures
for both protons and neutrons while this is not the case for
$^{100}$Sn, which may imply different core polarization effects on
$^{133}$Sb and $^{101}$Sn. 
We shall see that the softness of the core plays an important role in
the perturbative description of the single-particle structure of these
nuclei.

The two nuclei are described as a doubly-closed core plus one valence
nucleon, whose correlations with the core are taken into account by
means of the time-dependent degenerate linked-diagram perturbation
theory \cite{Kuo90}.
In previous works, single-particle valence nuclei of the $sd$-shell
have been extensively studied within the same framework by other
authors \cite{Kassis72,Maglione94,Sharon01}. 
However, the presence of the intruder state $0h_{11/2}$ makes a more
challenging problem to reproduce the structure of the
$2s1d0g0h$-shell. 

We have employed a class of onshell-equivalent realistic $NN$
potentials, derived from the CD-Bonn interaction \cite{Machleidt01b} by
way of the so-called $V_{\rm low-k}$ approach \cite{Bogner01}.
This approach is a very valuable tool since, by varying the  cutoff
momentum $\Lambda$, it provides different $NN$ potentials defined up
to $\Lambda$ which  are all onshell equivalent. This allows to study
how the short-range (high-momentum) components  of the $NN$ potential
contribute to determine the nuclear shell structure.

The paper is organized as follows. 
In Sec. II we give an outline of our calculations, focusing 
attention on the time-dependent degenerate linked-diagram expansion.
Sec. III is devoted to the presentation and discussion of our results
for the $^{133}$Sb and $^{101}$Sn single-particle spectra.
Some concluding remarks are given in Sec. IV.

\section{Outline of calculations}
As mentioned in the Introduction, we calculate the single-particle
energies of  $^{133}$Sb and $^{101}$Sn within the framework of the
time-dependent degenerate linked-diagram expansion \cite{Kuo71,Kuo90}.
This means that we describe the wave function of each nucleus as a
single-nucleon state $|j \rangle$, whose energy $\epsilon^j$ is
calculated taking into account perturbatively the interaction of the
odd nucleon with the closed core.
As a starting point, an auxiliary one-body potential $U$ is introduced
to break up the hamiltonian as the sum of an unperturbed term $H_0$,
which describes the independent motion of the nucleons, and a residual
interaction $H_1$:

\begin{eqnarray}
H & = & \sum_{i=1}^{A} \frac{p_i^2}{2m} + \sum_{i<j} V^{ij}_{NN} = T +
V_{NN} = \nonumber \\
~ & = & (T+U)+(V_{NN}-U)= H_{0}+H_{1}~~.
\label{smham}
\end{eqnarray}

The effective single-particle energies are expressed by way of the
folded-diagram expansion in terms of the so-called $\hat{Q}$ box:

\begin{equation}
\epsilon^j = \epsilon^{j}_{0} + \hat{Q} - \hat{Q}' \int \hat{Q} +
\hat{Q}' \int \hat{Q} \int \hat{Q} - \hat{Q}' \int \hat{Q} \int
\hat{Q} \int \hat{Q} + ~...~~,
\label{heff}
\end{equation}

\noindent
$\epsilon^{j}_{0}$ being the unperturbed energy of the single-particle
state $|j \rangle$, i.e.

\begin{equation}
H_{0} |j \rangle = (T+U) |j \rangle = \epsilon_{0}^{j} |j \rangle ~~.
\end{equation}

Eq. (\ref{heff}) represents the well known Kuo-Lee-Ratcliff (KLR)
folded-diagram expansion \cite{Kuo71,Kuo90}.
This expansion is written in terms of the vertex function
$\hat{Q}$-box, which is composed of one-body irreducible
valence-linked Goldstone diagrams. 
Once the $\hat{Q}$-box has been calculated through a certain order in
the input interaction $V_{NN}$, the series of the folded diagrams is summed
up to all orders using the Lee-Suzuki iteration method \cite{Suzuki80}.

\begin{figure}[H]
\begin{center}
\includegraphics[scale=0.6,angle=0]{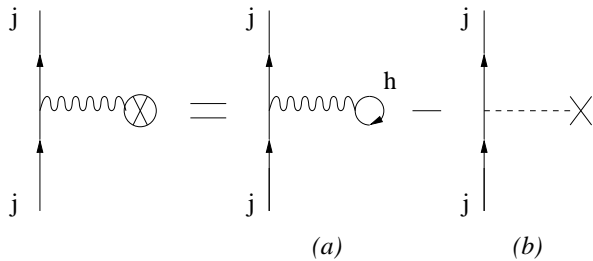}
\caption{First-order diagram. Graph ($a$) is the so-called self-energy
diagram. Graph ($b$) represents the matrix element of the harmonic
oscillator potential $U=\frac{1}{2}m\omega r^2$.}
\label{Sbox1}
\end{center}
\end{figure}

\begin{figure}[H]
\begin{center}
\includegraphics[scale=0.6,angle=0]{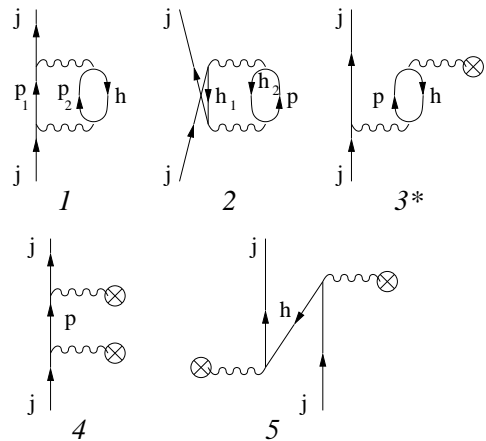}
\caption{Second-order diagrams. The asterisk indicates non-symmetric
 diagrams, which occur always in pairs giving equal contributions.
 For the sake of simplicity we report only one of them.}
\label{Sbox2}
\end{center}
\end{figure}

\begin{widetext}
\begin{figure}[H]
\begin{center}
\includegraphics[scale=0.9,angle=0]{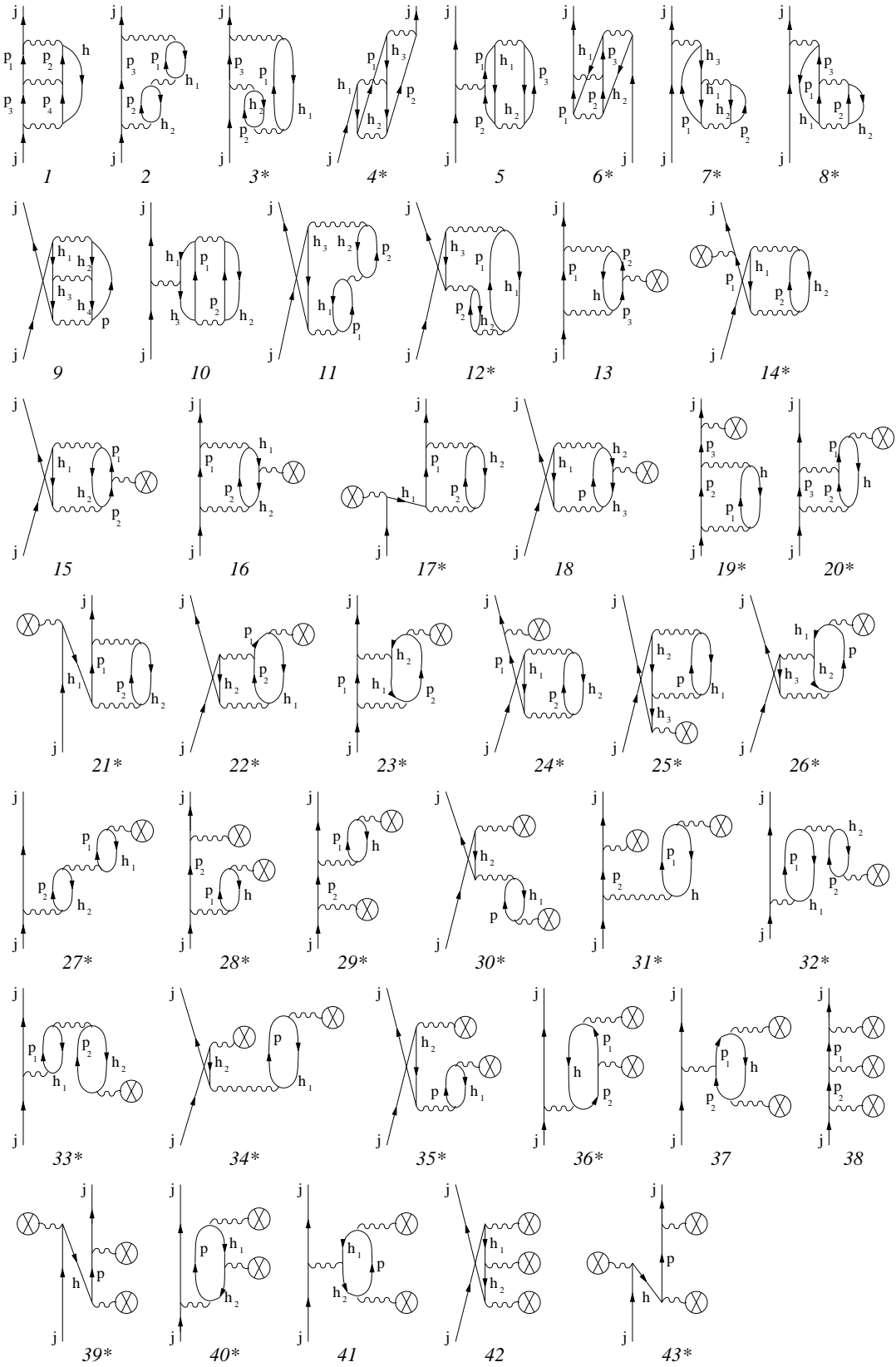}
\caption{Same as Fig. \ref{Sbox2}, but for third-order diagrams.}
\label{Sbox3}
\end{center}
\end{figure}
\end{widetext}

Our calculations have been performed including all the one-body
irreducible valence-linked Goldstone diagrams up to third order in
$V_{NN}$. 
In Fig. \ref{Sbox1} the first-order diagram, which is 
composed of the so-called self-energy diagram minus the auxiliary
potential $U$-insertion, is reported while the
second- and third-order diagrams are shown in
Figs. \ref{Sbox2} and \ref{Sbox3}, respectively. 
The $U$-insertion diagrams arise in the perturbative expansion owing
to the presence of the $-U$ term in $H_1$ (see Eq. (\ref{smham})).
In our calculation we have used as auxiliary potential the harmonic
oscillator (HO) one, which  is the most common choice.

Let us now focus our attention on the input interaction $V_{NN}$. 
Of course, it would be very  desirable to use directly a realistic
$NN$ potential that reproduces with high precision the two-body
scattering data and deuteron properties.
However, to perform nuclear structure calculations in a perturbative
framework with realistic $NN$ potentials one has first to deal with
the strong repulsive behavior of such potentials in the high-momentum 
regime. 
An advantageous method to renormalize the bare $NN$ interaction 
has been proposed in \cite{Bogner01,Bogner02}.
It consists in deriving from $V_{NN}$ a low-momentum potential 
$V_{\rm low-k}$ defined within a cutoff momentum $\Lambda$. 
This is a smooth potential which preserves exactly the onshell
properties of the original $V_{NN}$ and is suitable for being used 
directly in nuclear structure calculations. 

Let us now outline briefly the derivation of $V_{\rm low-k}$
\cite{Bogner02}.
The repulsive core contained in $V_{NN}$ is smoothed by integrating
out the high-momentum modes of $V_{NN}$ down to $\Lambda$. 
This integration is carried out with the requirement that the deuteron
binding energy and low-energy phase shifts of $V_{NN}$ are preserved
by $V_{\rm low-k}$. 
This is achieved  by the following $T$-matrix
equivalence approach. 
We start from the half-on-shell $T$ matrix for $V_{NN}$ 
\begin{eqnarray}
 T(k',k,k^2) =  V_{NN}(k',k) +
 ~~~~~~~~~~~~~~~~~~~~~~~~~~~~~~~~~~~~~~~~ \nonumber \\ 
\mathcal{P} \int _0 ^{\infty} q^2 dq V_{NN} (k',q)
\frac{1}{k^2-q^2} T(q,k,k^2 ) ~~, ~~~~~~~~~~~~~~~~~~~~~~~
\end{eqnarray}

\noindent
where $\mathcal{P}$ denotes the principal value and  $k,~k'$, and $q$
stand for the relative momenta. 
The effective low-momentum $T$ matrix is then defined by
\begin{eqnarray}
T_{\rm low-k } (p',p,p^2) =  V_{\rm low-k }(p',p) + 
~~~~~~~~~~~~~~~~~~~~~~~~~~~~~~~ \nonumber \\ 
\mathcal{P} \int _0 ^{\Lambda} q^2 dq  V_{\rm low-k }(p',q) 
\frac{1}{p^2-q^2} T_{\rm low-k} (q,p,p^2) ~~,~~~~~~~~~~~
\end{eqnarray}

\noindent
where the intermediate state momentum $q$ is integrated from 0 to the
momentum space cutoff $\Lambda$ and $(p',p) \leq \Lambda$. 
The above $T$ matrices are required to satisfy the condition 
\begin{equation}
T(p',p,p^2)= T_{\rm low-k }(p',p,p^2) \, ; ~~ (p',p) \leq \Lambda \,.
\end{equation}

The above equations define the effective low-momentum interaction 
$V_{\rm low-k}$, and it has been shown \cite{Bogner02} that their
solution is provided by the same KLR folded-diagram expansion
\cite{Kuo71,Kuo90} mentioned before.

\newpage
\newpage

In addition to the preservation of the half-on-shell $T$ matrix, which
implies preservation of the phase shifts, this $V_{\rm low-k}$
preserves the deuteron binding energy, since eigenvalues are preserved
by the KLR effective interaction. 
For any value of $\Lambda$, the $V_{\rm low-k}$ can be calculated very
accurately using iteration methods. 
Our calculation of $V_{\rm low-k}$ is performed by employing the
iteration method proposed in \cite{Andreozzi96}, which is based on the 
Lee-Suzuki similarity transformation \cite{Suzuki80}. 

We have derived from the high-precision CD-Bonn $NN$ potential
\cite{Machleidt01b} a class of $V_{\rm low-k}$'s corresponding to values
of the cutoff $\Lambda$ ranging from $2.1$ to $2.6$ fm$^{-1}$.
These $V_{\rm low-k}$'s are all onshell equivalent, more precisely
they all yield the same deuteron binding energy and $NN$ scattering
data up to the anelastic threshold, which corresponds to a relative
momentum $k \simeq 2.1$ fm$^{-1}$, as the CD-Bonn potential.
However, this does not imply that these $V_{\rm low-k}$'s  give
the same results when employed in the calculation of the
single-particle spectra, since the larger the value of the cutoff the
more the repulsive high-momentum components are explicitly included in
$V_{\rm low-k}$. 
In other words, $V_{\rm low-k}$'s corresponding to different cutoffs
have different offshell behaviors (see for istance \textcite{Coraggio05b}).
To illustrate this, let us consider the offshell tensor force
strength. 
This is related to the $D$-state probability of the deuteron $P_D$,
which implies that, when comparing $V_{\rm low-k}$'s with different
cutoffs, offshell differences are seen in $P_D$ differences.
From inspection of  Table \ref{pd}, where the $P_D$'s for various  
$V_{\rm low-k}$'s are reported, it can be seen that the larger the
value of the cutoff the stronger the offshell tensor force. 

\begin{table}[H]
\caption{Values of $P_D$ for $V_{\rm low-k}$'s corresponding to
  different cutoff momenta $\Lambda$ (in fm$^{-1}$) compared with the
  value yielded by the original CD-Bonn potential.}
\begin{ruledtabular}
\begin{tabular}{ccccccc}
 2.1 &  2.2 &  2.3 &  2.4 &  2.5 & 2.6 &  CD-Bonn \\
\colrule
3.96 & 4.09 & 4.21 & 4.32 & 4.41 & 4.49 & 4.85 \\
\end{tabular}
\end{ruledtabular}
\label{pd}
\end{table}

\section{Results}
\begin{figure}[H]
\begin{center}
\includegraphics[scale=0.4,angle=0]{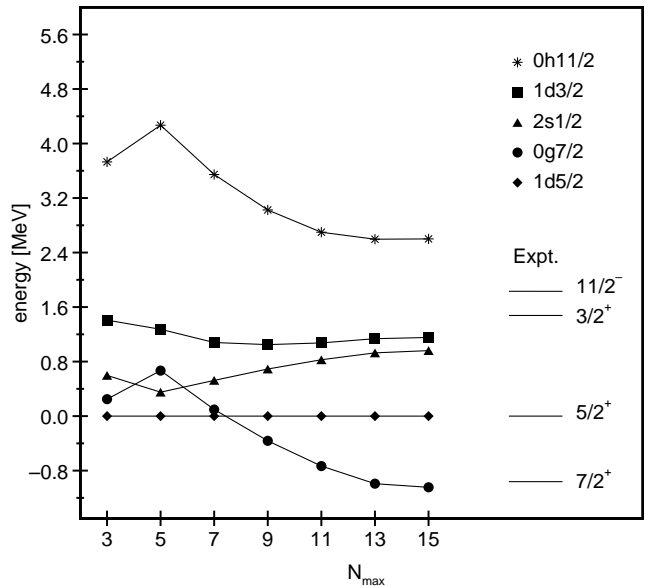}
\caption{Theoretical single-particle spacings in $^{133}$Sb relative
  to the $1d_{5/2}$ level, as a function of $N_{\rm max}$ (see text
  for details). The calculations refer to a cutoff $\Lambda=2.6$
  fm$^{-1}$. The experimental spectrum of $^{133}$Sb is also
  reported.} 
\label{Sboxconv}
\end{center}
\end{figure}

\begin{figure}[H]
\begin{center}
\includegraphics[scale=0.4,angle=0]{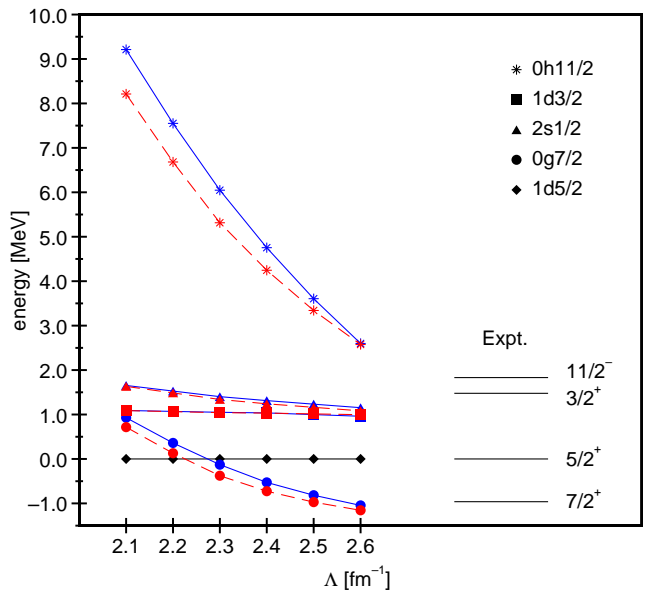}
\caption{(Color online) Calculated and experimental single-particle
  spacings in $^{133}$Sb relative to the $1d_{5/2}$ level (black
  line), as a function of the cutoff momentum $\Lambda$. The blue
  lines are results at third-order in the perturbative expansion. The
  red dashed lines refer to results obtained with the Pad\'e
  approximant $[2|1]$ (see text for details).}
\label{133Sb5}
\end{center}
\end{figure}

Calculated and experimental single-particle spacings for the one-proton
valence nucleus $^{133}$Sb are reported in Fig. \ref{133Sb5} as
a function of the cutoff $\Lambda$.
We have found it appropriate for the  perturbative expansion to limit
the value of $\Lambda$ to 2.6 fm$^{-1}$, since larger values of the
cutoff lead to a rapid deterioration of the the convergence properties.
The calculations of diagrams have been performed using intermediate
states whose excitation energy is up to $N_{\rm max}=15$
harmonic-oscillator quanta, so to obtain the stability the stability
of the results within few tens of keV when increasing the number of
intermediate particle states.
This is shown in Fig. \ref{Sboxconv}, where we report the relative SP
energies calculated at third order in $H_1$, with a cutoff value
$\Lambda=$2.6 fm$^{-1}$, as a function of $N_{\rm max}$.
The oscillator parameter used is $\hbar \omega = 7.88$ MeV, according
to the expression $\hbar \omega= 45 A^{-1/3} -25 
A^{-2/3}$ \cite{Blomqvist68} for $A=132$, and for protons the Coulomb
force has been explicitly added to the $V_{\rm low-k}$ potential.

It is important to point out that our model space is spanned by the
$0g_{7/2}$, $1d_{5/2}$, $1d_{3/2}$, $2s_{1/2}$, and $0h_{11/2}$ states
above the $^{132}$Sn core, so we have downshifted the unperturbed
energies of the $0h_{11/2}$ and $0g_{9/2}$ levels by a quantity equal
to $\hbar \omega$.
In this way the $2s1d0g0h$ shell is degenerate, as required by the
theory of the time-dependent degenerate linked-diagram expansion, and
the $0g_{9/2}$ acts as a hole state below the Fermi surface.
Anyway, the inclusion of diagrams with $(V-U)$-insertions in the
$\hat{Q}$ box assures  that our results are perturbatively independent
of the choice of $U$.
In fact, to sum up the $(V-U)$-insertion diagrams to all
orders is equivalent to employ a Hartree-Fock self-consistent
basis.

The results shown in Fig. \ref{133Sb5} (blue lines) are normalized
with respect to the $0d_{5/2}$ state (straight black line) in order to
evidence the changes in the spacings when increasing the cutoff
$\Lambda$. 
In this figure, we have also reported (red dashed lines) the results
obtained by calculating the Pad\'e approximant $[2|1]$ 
\begin{equation}
[ 2|1 ]=\hat{Q}_0+\hat{Q}_1+\frac{\hat{Q}_2}{1-\hat{Q}_{3}/\hat{Q}_2}~~,
\end{equation}

\noindent
$\hat{Q}_i$ being the the vertex function $\hat{Q}$-box calculated
through the $i$th-order in the input interaction $V_{NN}$.
Using Pad\'e approximants \cite{Baker70,Ayoub79} one may obtain an
estimate of the value to which the perturbative series should
converge.
So the comparison between the third-order results and those obtained
by means of the  Pad\'e approximant $[2|1]$ is an indicator of the
dependence of our results on the higher-order perturbative terms.

From inspection of Fig. \ref{133Sb5}, it can be seen that the
calculated spin-orbit splitting between $1d_{3/2}$ and $1d_{5/2}$
levels is almost independent from $\Lambda$, the experimental
separation being reasonably well reproduced. 
The distance  between the $2s_{1/2}$  energy and the centroid of
the $1d$ levels is rather independent from the cutoff too, but there
is no experimental counterpart.
It is worth to point out that the $1d_{3/2}$ and $2s_{1/2}$ relative
energies do not exhibit a significant dependence on higher-order
contributions, the difference between the third-order results and
those obtained with the $[2|1]$ Pad\'e approximant being very small. 

The situation is very different for the calculated relative energies
of the $0g_{7/2}$ and $0h_{11/2}$ levels, whose spin-orbit partners
have been  placed outside the chosen model space.
The calculated $0g_{7/2}$, $0h_{11/2}$ relative energies show a strong
dependence on the cutoff $\Lambda$ and a slightly worse perturbative
behavior.
The most striking feature is that for small values of the cutoff the
discrepancies with respect to the experimental data are quite large,
while a better agreement is obtained increasing $\Lambda$. 
In particular, it should be pointed out that with small values of the
cutoff the $0h_{11/2}$ level is far away in energy from the other
levels, thus implying a shell closure at $Z=70$. 
A much more reasonable spectrum of $^{133}$Sb is obtained when
increasing the value of the cutoff which cause the intruder
$0h_{11/2}$ proton state to join  the 50-82 shell. 
This evidences the non-negligible role of the high-momentum components
of $V_{NN}$ in the formation of the observed shell structure. 

However, after these remarks one may wonder whether when increasing
$\Lambda$ the model space should be enlarged to include the $2p1f0h$
levels.
More precisely, for large value of the cutoff it might occur a sort of
``shell-quenching'' when calculating the single-particle spectrum
above the closed-shell $^{132}$Sn.
As a consequence, the $0h_{11/2}$ state as well as its parity partners
($0h_{9/2},~1f_{7/2},~1f_{5/2},~2p_{3/2},~2p_{1/2}$) would approach in
energy the $2s1d0g$ states,  leading to the birth of a sort of
``macro''-shell $2s1d0g2p1f0h$.

\begin{figure}[H]
\begin{center}
\includegraphics[scale=0.4,angle=0]{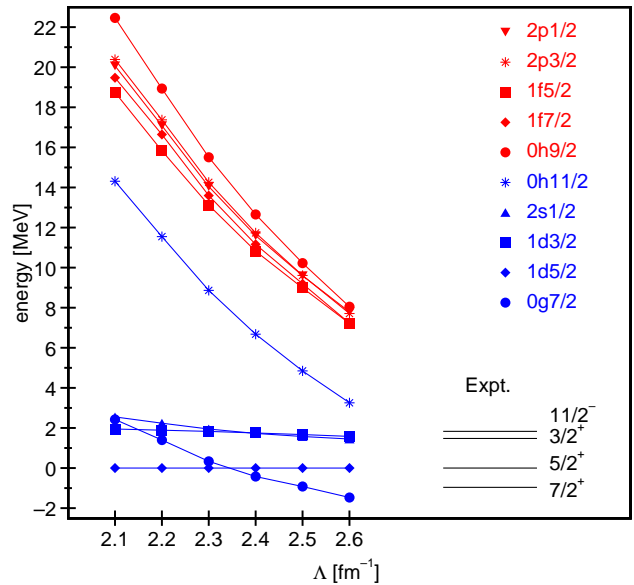}
\caption{(Color online) Calculated and experimental single-particle
  spacings in $^{133}$Sb relative to the $1d_{5/2}$ level (black
  line), as a function of the cutoff $\Lambda$. The blue lines denote
  the $2s1d0g0h$ levels, while the red one the $2p1f0h$ levels (see
  text for details).}
\label{133Sb10}
\end{center}
\end{figure}

In order to investigate this question, we have performed a
kind of theoretical ``experiment''.
We have carried out calculations where the $2p1f0h$ proton-particle
unperturbed energies have been considered degenerate with the $2s1d0g$
ones, and then we have turned on the interaction $H_1$.
The calculated energies relative to the $0d_{5/2}$ level are
reported in Fig. \ref{133Sb10}.
It can be seen that the spin-orbit splittings of the $2p1f0h$ levels
are practically independent of the cutoff $\Lambda$, while the
$0h_{11/2}$ level is pushed down from its parity partners.
In particular, for large value of the cutoff the $0h_{11/2}$ state is
much closer in energy to the $2s1d0g$ levels, confirming the
reliability of a model space spanned only by the five $2s1d0g0h$
states.

We have found it interesting to investigate the counterpart of the
$2s1d0g0h$ shell for neutrons, namely the single-particle spectrum of
$^{101}$Sn.
In such a case, the closed-shell core is the exotic $N=Z$ nucleus
$^{100}$Sn. 
To this end, we have performed the same  calculation as for
$^{133}$Sb, except that the harmonic-oscillator parameter is $\hbar
\omega = 8.55$  MeV.

\begin{figure}[H]
\begin{center}
\includegraphics[scale=0.4,angle=0]{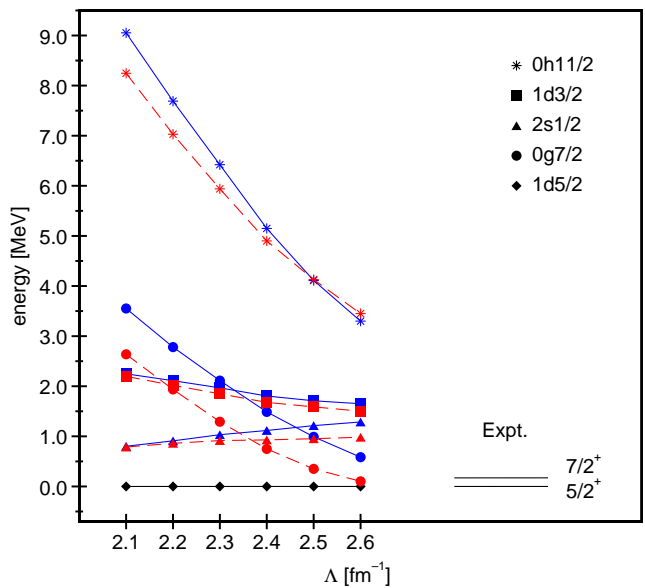}
\caption{(Color online) Same as in Fig. \ref{133Sb5}, but for $^{101}$Sn}
\label{101Sn5}
\end{center}
\end{figure}

In Fig. \ref{101Sn5} we report the calculated and experimental
single-neutron energies for $^{101}$Sn relative to the $1d_{5/2}$
state as a function of the cutoff $\Lambda$.
As can be seen, the main difference with respect to the results shown
in Fig. \ref{133Sb5} relies in the convergence properties of the
$0g_{7/2}-1d_{5/2}$ spacing, that is the only spacing observed in
$^{101}$Sn.
More precisely, for the $0g_{7/2}$ energy level in $^{101}$Sn we have
found that the third-order contributions of diagrams (27), (32), and
(33) of Fig. \ref{Sbox3}, with respect to the second order diagram (3)
of Fig. \ref{Sbox2}, are larger than what has been found for
$^{133}$Sb.
All these graphs represent the so-called Tamm-Dancoff plus
Random-Phase Approximation diagrams.

The stronger role of higher-order perturbative terms in the
linked-diagram expansion is probably due to the fact that $^{100}$Sn
is a much ``softer'' core than $^{132}$Sn.

Apart these concernings about the perturbative behavior of the
linked-diagram expansion, our results confirm the need of the
high-momentum components of the nuclear force in order to improve the
description of the neutron shell structure.

\section{Summary and conclusions}
In this paper, we have calculated  the single-particle spectra of
$^{133}$Sb and $^{101}$Sn within the framework of the time-dependent
degenerate linked-diagram perturbation theory.
To investigate how the high-momentum components of the $NN$ potential
influence the structure of the  proton/neutron-$2s1d0g0h$ shell,
we have employed a class of onshell-equivalent realistic
nucleon-nucleon potentials.
These potentials have been derived from the high-precision CD-Bonn
interaction \cite{Machleidt01b} by way of the so-called 
$V_{\rm low-k}$ approach, varying the cutoff momentum from 2.1 up to
2.6 fm$^{-1}$.

The perturbative behavior of the calculated single-particle
relative energies is quite good  for $^{133}$Sb, while is less
satisfactory for $^{101}$Sn.
This may be traced  to the higher degree of exoticity of $^{101}$Sn,
which is very  close to the proton drip line. 

Our results show that, while the calculated spin-orbit splittings are
scarcely dependent on the cutoff value, the relative energy of the
intruder $0h_{11/2}$ state is strongly sensitive to the explicit
inclusion of the high-momentum components of the $NN$ potential.
We have found that for small values of the cutoff the $0h_{11/2}$
level lies far away in energy from the other single-particle levels,
while the correct shell structure is restored when increasing the
cutoff.

However, it should be noted that the lack of the high-momentum
components could be compensated by the inclusion of an effective
three-body force \cite{Schwenk06,Hagen07}, which, for small cutoffs,
may play a non-negligible role in determining the single-particle
spectrum \cite{Coraggio07b}.

\begin{acknowledgments}
This work was supported in part by the Italian Ministero
dell'Istruzione, dell'Universit\`a e della Ricerca  (MIUR), by the
U.S. DOE Grant No. DE-FG02-88ER40388.
\end{acknowledgments}

\bibliographystyle{apsrev}
\bibliography{biblio}

\newpage

\newpage

\newpage

\newpage

\newpage

\newpage


\end{document}